\documentclass[aps,english,prl,twocolumn,superscriptaddress]{revtex4}
\usepackage{graphicx}

\begin{document}

\title{Stable single-layer honeycomb like structure of silica}

\author{V. Ongun \"{O}z\c{c}elik}
\affiliation{UNAM-National Nanotechnology Research Center, Bilkent University, 06800 Ankara, Turkey}
\affiliation{Institute of Materials Science and Nanotechnology, Bilkent University, Ankara 06800, Turkey}
\author{S. Cahangirov}
\affiliation{Nano-Bio Spectroscopy Group, Dpto. Fisica de Materiales, Universidad del Pais Vasco, Centro de Fisica de Materiales CSIC-UPV/EHU-MPC and DIPC, Av. Tolosa 72, E-20018 San Sebastian, Spain}
\affiliation{Department of Physics, Bilkent University, Ankara 06800, Turkey}
\author{S. Ciraci}
\affiliation{Department of Physics, Bilkent University, Ankara 06800, Turkey}

\begin{abstract}
Silica or SiO$_2$, the main constituent of earth's rocks has several 3D complex crystalline and amorphous phases, but it does not have a graphite like layered structure in 3D. Our theoretical analysis and numerical calculations from the first-principles predict a single-layer honeycomb like allotrope, h$\alpha$-silica, which can be viewed to be derived from the oxidation of silicene and it has intriguing atomic structure with re-entrant bond angles in hexagons. It is a wide band gap semiconductor, which attains remarkable electromechanical properties showing geometrical changes under external electric field. In particular, it is an auxetic metamaterial with negative Poisson's ratio and has a high piezoelectric coefficient. While it can form stable bilayer and multilayer structures, its nanoribbons can show metallic or semiconducting behavior depending on their chirality. Coverage of dangling Si orbitals by foreign adatoms can attribute new functionalities to h$\alpha$-silica. In particular, Si$_2$O$_5$, where Si atoms are saturated by oxygen atoms from top and bottom sides alternatingly can undergo a structural transformation to make silicatene, another stable, single layer structure of silica.
\end{abstract}

\maketitle

Trends in materials science have aimed at the discovery of single layer structures and their multilayer van der Waals composites by using advanced fabrication techniques \cite{search}. In this endeavor, silicene, a graphene like single layer honeycomb structure of silicon, was shown to be stable \cite{seymur1,stability} and later synthesized on Ag(111) substrate \cite{lelay1}. The interaction of silicene with oxygen has been a subject of active research \cite{weiss2005,loffler2010,padova,ongun1,rwang,molle}, because it directly affects the fabrication process of silicene-based devices. Besides this, silicene-oxygen interaction leads to new two-dimensional(2D) silica crystals in different forms \cite{weiss2005, loffler2010} and bilayer structures \cite{huang2012}. Recently, efforts have been devoted to grow 2D ultrathin polymorphs of silica on substrates \cite{todorova,winter,huang2012,loffler2010,freund} and theoretical studies have been also carried out to understand 2D silica and their defects \cite{rwang,bjorkman,durandurdu}. In various 3D fourfold coordinated allotropes of SiO$_2$, such as amorphous and crystalline quartz, which are commonly named as silica, Si-O-Si bonds are bent around oxygen atoms \cite{parinello}, except for $\beta$-cristobalite which has straight Si-O-Si bonds \cite{salim}. However, none of those allotropes of silica are known to have a graphene like freestanding 2D structure or a graphite like layered structure.

\begin{figure}
\includegraphics[width=6cm]{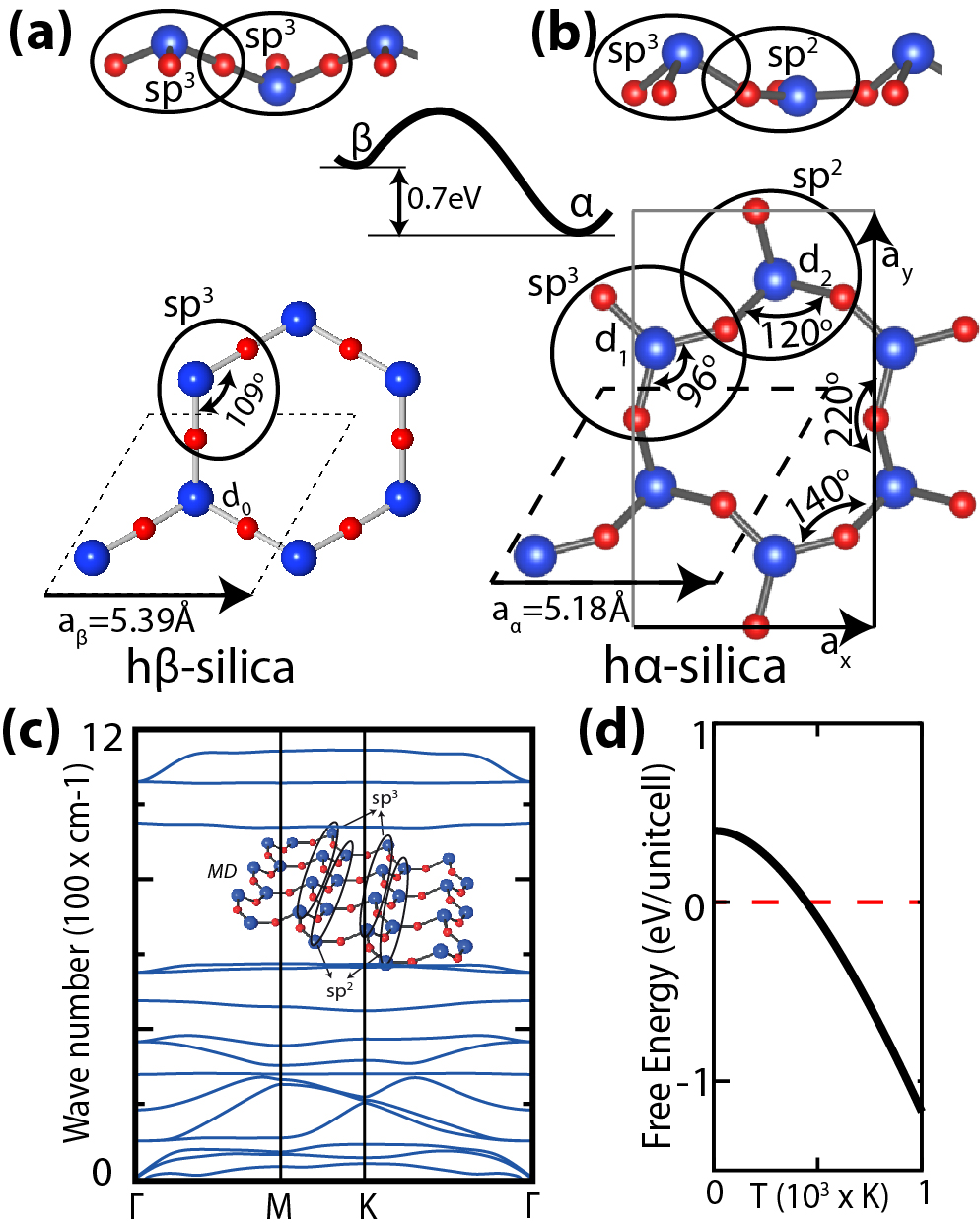}
\caption{(a) Side and top views of h$\beta$-silica. Large blue and small red balls stand for Si and O atoms. (b) Optimized structure of h$\alpha$-silica with its hexagonal and rectangular unit cells. Two types of Si atoms, i.e. $sp^3$-bonded and $sp^2$-bonded are indicated. (c) Calculated phonon bands and tilted view of the MD simulation in $4 \times 4$ unitcell. (d) Free energy versus temperature.}
\label{fig1}
\end{figure}

\begin{figure*}
\includegraphics[width=12cm]{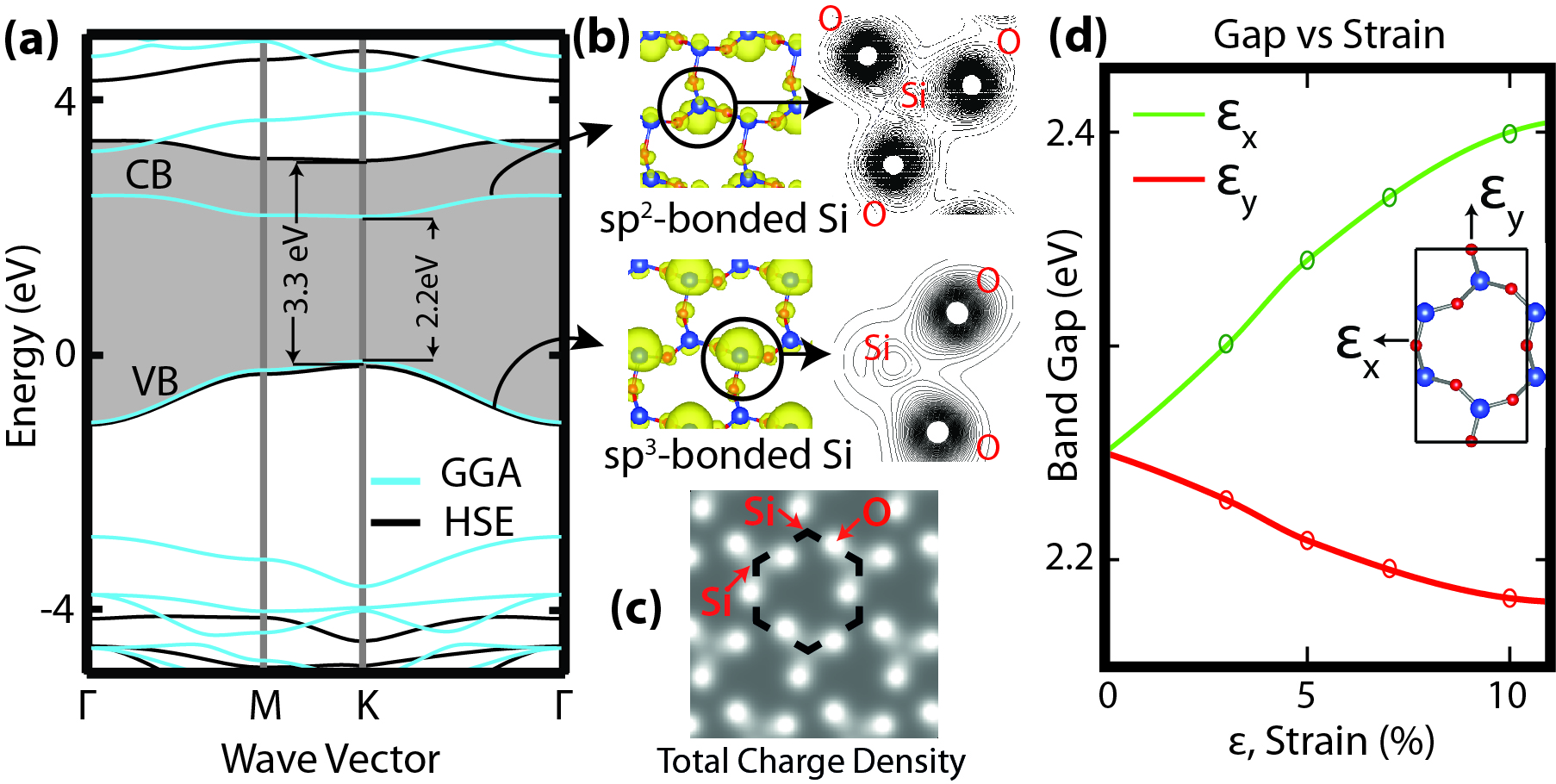}
\caption{ (a) The electronic band structure of h$\alpha$-silica. (b) Isosurface charge densities of the lowest conduction(CB) and highest valence(VB) band, and contour plots of the total charge density in the planes of O-Si-O for the $sp^2$ and the $sp^3$-bonded Si atoms. (c) Images of the total charge density. (d) Variation of the band gap as a function of the uniaxial strains $\epsilon_x$ and $\epsilon_y$.}
\label{fig2}
\end{figure*}

In this letter, we predict a single layer allotrope of silica named as h$\alpha$-silica (h denoting the single layer hexagonal lattice). Having passed through various complex stability tests, this honeycomb like structure is found to be stable. It has more complex structure and exceptional properties as compared to those ultrathin films grown on substrates; it is a wide band gap semiconductor having a negative Poisson's ratio and high piezoelectric coefficient. The band gap can be reduced or widened depending on the direction of the applied uniaxial tensile stress. Even more remarkable, this basic structure, h$\alpha$-silica can form also bi-layer and multilayer structures, and interact readily with adatoms like O, H, F to form a manifold of its derivatives. In particular, when Si atoms are saturated by oxygen, the resulting derivative Si$_2$O$_5$, is transformed to silicatene at elevated temperature. Our predictions are obtained from first-principles calculations, which are detailed in Ref[\onlinecite{method}].

Silicene surface is rather reactive; O atoms are bonded to the bridge site with a strong binding energy of 6.17 eV \cite{ongun1}. This bridge bonded O adatom can forcibly switch between two bistable equilibrium positions at either sides if an energy barrier of 0.28 eV is overcame \cite{ongun1}. At the transition state, the Si-Si bond is stretched to accommodate one O atom near its midpoint and hence to form a Si-O-Si bond. Then, one may  contemplate whether an exothermic process can take place, in which one O atom can be inserted between each Si-Si bond of silicene concomitantly to grow a single layer honeycomb like network in 2D with the formula unit of Si$_2$O$_3$. Actually, similar process occurs in 3D through the oxidation of silicon surfaces to generate amorphous SiO$_2$.

A hexagonal unitcell including two Si and three O atoms forms a regular honeycomb network, as shown in Fig.~\ref{fig1}(a). This structure, which we named as h$\beta$-silica in analogy to $\beta$-cristobalite in 3D, corresponds to a shallow local minimum. When perturbed from perfect symmetry and subsequently relaxed by conjugate gradient method using stringent convergence criteria, h$\beta$-silica eventually transforms to a new structure, h$\alpha$-silica, and the energy is lowered by 0.7 eV. The optimized geometrical parameters of this allotrope and other physical properties are described in Fig.~\ref{fig1}(b) and Table~\ref{table1}. In the optimized structure, three alternating Si atoms of a hexagon engage in planar $sp^2$-bonding and the remaining three rise upwards or downwards \cite{updown} and form $sp^3$-like bonds \cite{sp3}. Consequently, the rotary reflection symmetry is broken and the lattice is shrank by $\sim$ 4\%. The straight Si-O-Si bonds are bent forming re-entrant bond angles \cite{reentrant}, such that all O atoms are coplanar together with  the $sp^{2}$-bonded Si atoms. Thus the regular hexagons are distorted, where Si atoms are placed at the corners of hexagons, but they are buckled like silicene. Its high cohesive energy, $E_c$=28.6 eV per unit cell originates from mixed (covalent and ionic) bonds  between Si and O atoms while its high formation energy prevents it from clustering. These rearrangements are reminiscent of the transition from ideal $\beta$-cristobalite to $\alpha$-quartz in 3D.

\begin{figure*}
\includegraphics[width=12cm]{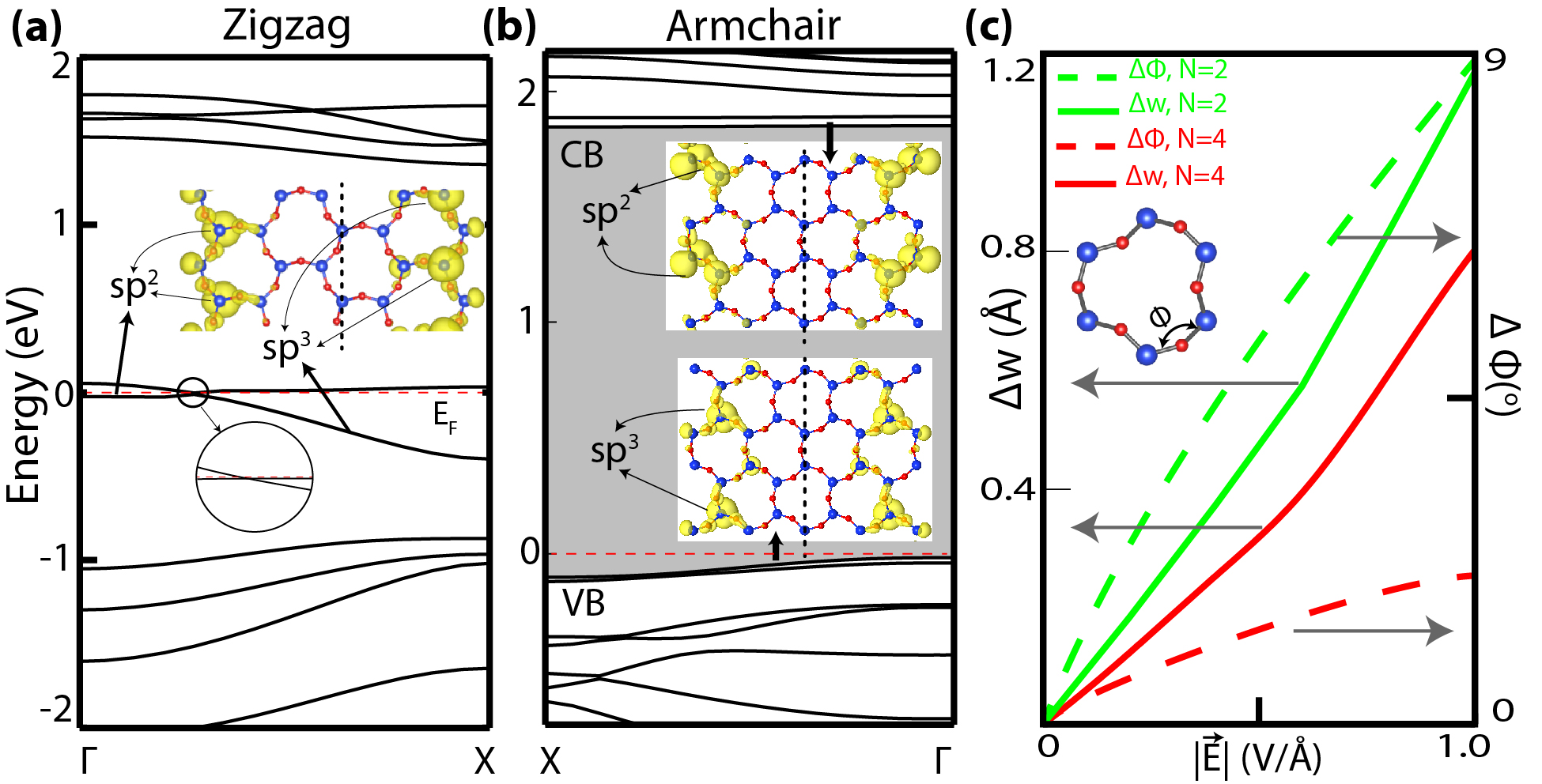}
\caption{(a) Electronic band structure of zigzag nanoribbon with $N_Z=4$. (b) Same for the armchair nanoribbon with $N_A=4$. (c) Variation of the width of the armchair nanoribbon $\Delta w$ and bond angle $\Delta{\phi}$ under an in-plane electric field $\vec{E}$ applied perpendicular to the axis.}
\label{fig3}
\end{figure*}

\begin{table*}
\caption{Relevant physical properties of h$\alpha$-silica. $a_{\alpha}$: hexagonal lattice constant in \AA; $d_1$: Si-O bond distance of $sp^3$-bonded Si; $d_2$: same for $sp^2$-bonded Si; $t$: perpendicular distance between $sp^{3}$-bonded and $sp^{2}$-bonded Si atoms in \AA; $E_{c}$, $E_f$: the cohesive and formation energies per unit cell in eV; $E_{G-gga}$: band gap calculated by GGA with van der Waals correction in eV; $E_{G-hse}$: band gap calculated by HSE;  $C$: in-plane stiffness in N/m; $\nu$: Poisson ratio; $Q^{*}_O$, $Q^{*}_{Si-sp^3}$, $Q^{*}_{Si-sp^2}$: Mulliken charges in electrons for different atoms.}
\label{table1}
\tabcolsep=0.1cm
\small
\begin{center}
\begin{tabular}{ccccccccccccccccc}
\hline  \hline
$a_{\alpha}$ & $d_1$ & $d_2$ & t & $E_{c}$ & $E_{f}$ & $E_{G-gga}$ & $E_{G-hse}$ & C & $\nu$ & $Q^{*}_O$ & $Q^{*}_{Si-sp^3}$ & $Q^{*}_{Si-sp^2}$\\
\hline
$5.18$ & $1.76$ & $1.58$ & 0.96 & $28.6$ & $9.2$ & $2.2$ & $3.3$ & 22.6 & -0.21 & $-0.31$ & $0.53$ & $0.39$  \\
\hline
\hline
\end{tabular}
\end{center}
\end{table*}

The rigorous answer to the question of whether the optimized structure is stable is provided by calculating the phonon frequencies with extreme accuracy. All frequencies of the infinitely large, single layer are found to be positive in the first Brillouin zone. As shown in Fig.~\ref{fig1}(c), the calculated bands of frequencies, $\Omega (\textbf{k})$, with a minute gap between optical and acoustical branches, clearly demonstrate that h$\alpha$-silica remains stable even if the stabilizing finite size and edge effects are absent. Otherwise, the specific eigenfrequencies of the dynamical matrix would be imaginary, if any instability due to the long wavelength transversal acoustical phonons were imposed.

Although the phonon spectrum provides supports that h$\alpha$-silica is a local energy minimum, we further examine the thermodynamic stability by calculating the variation of free energy with temperature. Our results that are derived from the phonon calculations including the entropy part indicate in Fig.~\ref{fig1}{d} that the system has positive free energy up to 500K.

To further investigate the effects of temperature on stability, we perform ab-initio molecular dynamics (MD) calculations at high temperatures. Since small unit cell may easily lead to fake instability, we used relatively large (4$\times$ 4) supercell. Starting from regular h$\beta$-silica, and by raising the temperature to 1000 K; the structure has transformed to h$\alpha$-silica and did not change in the course of MD simulations lasting 12 ps as shown by the inset of Fig.~\ref{fig1}(c).  No instability leading to structural transformation or dissociation of h$\alpha$-silica was observed even if the specific phonon modes were softened at such a high temperature. Additionally, through intermediate energy paths specific atoms were displaced forcibly from their optimized local equilibrium positions corresponding to a minimum in Born-Oppenheimer surface. The optimized structure was resistant to these deformations and hence restored itself quickly. This situation demonstrates that the minimum is deep enough to retain the structure of h$\alpha$-silica stable at ambient conditions.

Our conclusions on the mechanical stability is corroborated further by calculating its in-plane stiffness, $C= A_{o}^{-1}{\partial^{2} E_{T}/\partial\epsilon^{2}}$ (in terms of the area $A_o$ of the cell and the variation of total energy, $E_T$ with strain, $\epsilon$). Here calculations can conveniently be performed in rectangular cell by applying uniaxial strain along $x$- and $y$-directions to deduce that $C_x \simeq C_y =$22.6 $J/m^2$. The calculated in-plane stiffness is smaller than that of graphene \cite{ansikl}.

The Poisson's ratio $\nu=-\epsilon_{y}/\epsilon_{x}$, is calculated as $\nu=-0.21$. This is remarkable in the sense that as h$\alpha$-silica is stretched along $x$-direction it also expands in the $y$-direction, owing to its squeezed structure consisting of twisted and bent Si-O-Si bonds. Thus, three oxygen atoms in each hexagon protrude inwards resulting in a re-entrant structure.\cite{reentrant} The negative Poisson's ratio is a rare situation \cite{poisson} and is also observed in foams produced from low density, open-cell polymers \cite{foam,xfwang}. These extremal materials, also called as auxetics or metamaterials, offer crucial applications for biotechnology and nanosensors.

Electronic properties of h$\alpha$-silica are summarized in Fig.~\ref{fig2} which shows that it is nonmagnetic, wide band gap semiconductor with a direct band gap of $E_{G-gga}$=2.2 eV predicted using GGA calculations \cite{pbe}. When corrected by HSE \cite{hse}, the gap increases to $E_{G-hse}$=3.3 eV. While conduction and valance band edges are mainly derived from the dangling orbitals of $sp^2$-bonded and $sp^3$-bonded Si, respectively. The bands of oxygen-$p$ orbitals occur below $\sim$-4 eV.

According to Mulliken charge analysis \cite{siesta} summarized in Fig.~\ref{fig2}(b), Si-O bonds have a strong ionic character. Due to the special bond configuration, each oxygen atoms receives $\sim$0.31 excess electrons, while 0.53 electrons and 0.39 electrons are donated by $sp^3$- and $sp^2$-bonded Si atoms, respectively. The details of charge transfer are depicted by charge density isosurfaces and contour plots. Charge rearrangements are also reflected by the calculated total charge density in Fig.~\ref{fig2}(c), where oxygen and corner Si atoms are seen by bright and dispersed spots, respectively, while $sp^3$-bonded Si atoms are not seen. Due to the significant charge transfer between Si and O atoms, dipole moments along $x$- and $y$-directions are expected to be high.

It is a rather rare situation that the variation of the calculated band gap is strain specific; it increases with increasing uniaxial strain, $\epsilon_x$; but it decreases with increasing $\epsilon_y$ as shown in Fig.~\ref{fig2}(d). Reversed responses of the band gap to the orthogonal uniaxial tensile stresses are closely related to the atomic configuration in the rectangular unit cell described in Fig.~\ref{fig1}.

Nanoribbons attribute further relevance to h$\alpha$-silica. In Fig.~\ref{fig3} we consider zigzag and armchair nanoribbons specified by the number of hexagons across their widths $w$,  $N_Z$=4 and $N_A$=4, respectively. Apart from the minute reconstruction at their edges, these nanoribbons are stable. The zigzag $N_{Z}$=4  nanoribbon is a metal, where the flat band at the Fermi level is derived from the orbitals of $sp^2$-bonded atoms located at one edge. Another band crossing the Fermi level has small dispersion and originates from the orbitals of $sp^3$-bonded Si atoms at the other edge. In contrast, the armchair nanoribbon is a semiconductor with a direct band gap of 1.9 eV calculated within GGA \cite{pbe}. The bands at the conduction and valance band edges are derived from the $sp^2$-bonded and $sp^3$-bonded Si atoms at both edges, respectively. Due to small coupling between edges the bands are slightly split. These nanoribbons with or without foreign adatoms attached to the edges and their networks consisting of the combination of zigzag and armchair nanoribbons with different $N_Z$ and $N_A$, can display diversity of electronic properties, which is beyond the scope of the present study. Here we point out a property of h$\alpha$-silica, which may be of potential importance. Due to the significant charge transfer between Si and O atoms and its re-entrant structure; a finite size flake of h$\alpha$-silica is expected to be influenced by the applied electric field. The effect of the in-plane electric field, $\vec{E}$, is investigated on the armchair nanoribbons. As illustrated in Fig.~\ref{fig3}(c), applied in-plane electric field induces significant changes in the width $\Delta w$ and Si-O-Si bond angle $\phi$ of armchair nanoribbons. The piezoelectric coefficient is estimated from $\Delta w$ as 5.7x10$^{-12}$ m/V, which is more than twice the value measured for quartz \cite{virgil}, which is crucial for piezoelectric nanodevices.

\begin{figure}
\includegraphics[width=8cm]{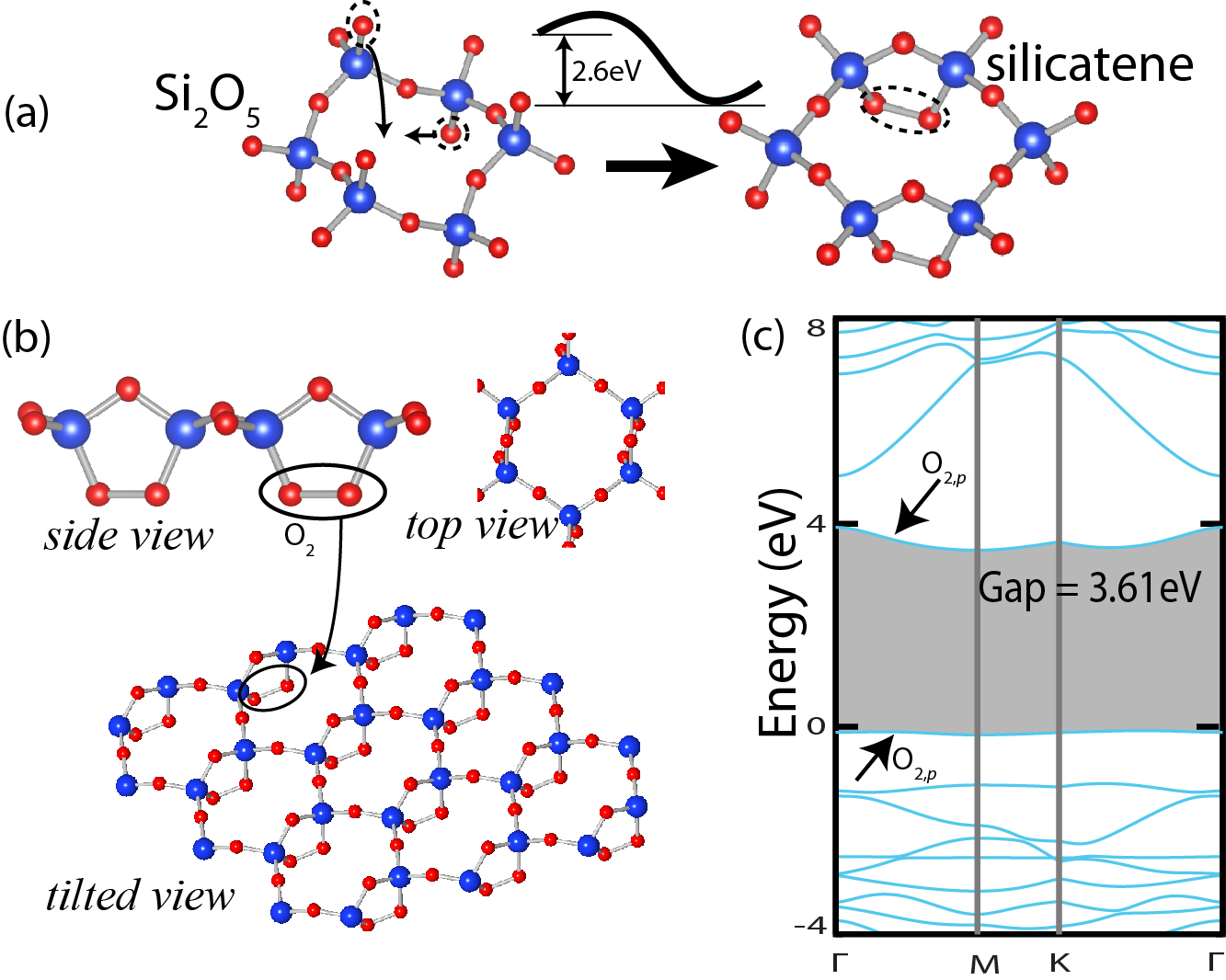}
\caption{(a) Silicatene derived from Si$_2$O$_5$. (b) Atomic structure of silicatene with side, top and tilted views. Pentagons are formed from three O and two Si atoms. (c) Electronic band structure. Bands derived from O-O bonds are indicated.}

\label{fig4}
\end{figure}

It should be noted that the basic structure h$\alpha$-silica and its ribbons discussed so far are reactive due to dangling bonds or $\pi$-orbitals oozing from Si atoms. Thus, they can acquire new functionalities by the adsorption of adatoms to Si, such as H, F, O, etc. Like graphene, uniform and full coverage of h$\alpha$-silica by one of these atoms can produce new derivatives Si$_2$O$_3$H$_2$, Si$_2$O$_3$F$_2$ and Si$_2$O$_5$. A summary of the properties attained by these derivatives,as well as bilayer and multilayer structures are presented as Supplementary Materials. Here the derivative Si$_2$O$_5$ is of particular interest. Upon the oxidation through the saturation of Si dangling bonds alternatingly from the top and bottom sides, the filled band at the top of the valance band derived from the orbitals of $sp^{3}$-bonded Si and the empty band at the bottom of the conduction band derived from the orbitals of $sp^{2}$-bonded Si atoms in Fig.~\ref{fig2} are removed and the band gap increases from 2.2 eV to $\sim$6 eV attributing a high insulating character and inertness like 3D silica. While the hexagon like 2D geometry in Fig.~\ref{fig1} is maintained, $sp^{2}$-bonded Si atoms change to $sp^{3}$-bonded Si atoms and hence restore the rotary reflection symmetry. This way, Si atoms acquire the fourfold coordination of oxygen atoms in Fig.~\ref{fig4} (a) as in 3D silica. However, upon heating Si$_2$O$_5$ undergoes a structural transformation by further lowering (i.e. becoming more energetic) its total energy by 2.63 eV, whereby first half of dangling Si-O bonds rotate from top to bottom so that all are relocated at the bottom side. Eventually, they are paired to form O-O bonds as shown in Fig. \ref{fig4} (a). The resulting electronic structure of this phase is presented in Fig.~\ref{fig4} (b). The filled and empty bands occurring in the band gap of Si$_2$O$_5$ is derived from O-O bonds. As a final remark, it is gratifying that the optimized atomic structure predicted in Fig. \ref{fig4} (a) replicates the structure of the single layer silica in honeycomb structure (named as silicatene), the growth of which was achieved much recently on Ru(0001) surface \cite{yang}. The measured lattice parameters also agrees with the present calculated values. Based on the present analysis it is revealed that silicatene is bonded to metal substrates through O-O bonds below the Si plane.

In conclusion, we showed that the single layer honeycomb like structure of silica h$\alpha$-silica is stable and mimics the well-known crystalline quartz in 2D. Additionally, this allotrope and its nanoribbons display exceptional electromechanical properties and can attain bilayer and multilayers through the stacking of silica single layers. While diverse functionalities can be attained by the adsorption of foreign atoms, a new phase named as silicatene forms when the dangling Si orbitals are saturated by oxygen atoms. While the allotropes of bulk silica have made great impact in fundamental and applied research, the exceptional properties revealed in this paper herald that h$\alpha$-silica and its derivatives can present potential applications, such as composite materials, functional coating and also as single layer insulator in nanoelectronics and nanocapacitors.


\begin{thebibliography}{15}
\bibitem{stability}
The fact that the free standing graphene does exist, but not silicene, cannot be taken as a counter argument against the stability of silicene. All single layer structures including graphene need a suitable substrate if they are growing from constituent atoms. Free standing silicene does not exist, since there is no 3D allotrope of silicon consisting of weakly bonded layers like graphite, BN or MoS$_2$
\bibitem{updown}
From the symmetry point of view it is equally probable that $sp^3$-bonded Si atoms can be either at the top or bottom side of the plane of oxygen atoms. This implies frustration and may lead to domain structure.
\bibitem{sp3}
Even though the bond angle of $sp^3$-hybrid bonds deviate from tetrahedral angles, we continue to specify related Si atoms as $sp^3$-bonded Si atoms throughout the text.
\bibitem{reentrant}
The Si-O-Si angles between $sp^2$- and $sp^3$-bonded Si atoms, which are denoted by $\phi$ for oxygens protruding outwards and by $\phi^{'}$ for oxygen atoms protruding inwards, are 140$^o$ and 220$^{o}$, respectively.
\bibitem{poisson}
In Ref[\onlinecite{ansikl}], the calculated Poisson's ratio of more than 20 stable, single layer honeycomb structure of Group IV elements and Group III-V compounds were found to be positive.
\end{thebibliography}
\end{document}